\documentclass{appolb}
\usepackage{graphicx}

\newcommand{\dsp}{\displaystyle}
\newcommand{\ba}{\begin{eqnarray}}
\newcommand{\ea}{\end{eqnarray}}
\newcommand{\be}{\begin{equation}}
\newcommand{\ee}{\end{equation}}
\begin{document}

\begin{titlepage}
  \begin{flushright}
    LU TP 07-03\\
NT@UW-07-02\\
    hep-ph/0701267\\
    January 2007
  \end{flushright}
\vskip2cm
\begin{center}
  {\Large\bf Partially Quenched and Three Flavour ChPT \\
at Two Loops%
\footnote{Presented by JB at ``Final Euridice Meeting'', August 24-28  2006, Kazimierz, Poland.}}

\vskip1.5cm

{\bf Johan Bijnens$^{a}$, Niclas Danielsson$^{a,b}$ and
Timo A. L\"ahde$^{c}$}\\[1cm]

{$^a$Department of Theoretical Physics, Lund University,\\
S\"olvegatan 14A, SE 22362 Lund, Sweden\\[4mm]
$^b$Division of Mathematical Physics,
Lund University,\\
S\"olvegatan 14A, SE 22362 Lund, Sweden\\[4mm]
$^c$Department of Physics,
University of Washington,\\
Seattle, WA 98195-1560,
USA}
\end{center}

\vskip2cm
\begin{abstract}
A summary of recent progress in Chiral Perturbation Theory (ChPT) at the two-loop 
level is given. A short introduction to ChPT is included, along with an explanation 
of the usefulness of developing ChPT for partially quenched QCD. Further, our recent 
work in partially quenched ChPT is reviewed, and a few comments are given on older 
work in mesonic ChPT at the two-loop level. In particular, we quote the present 
best values for the low-energy constants of the $\mathcal{O}(p^4)$ chiral Lagrangian. 
\end{abstract}

\end{titlepage}
\setcounter{page}{1}
\setcounter{footnote}{0}

\title{Partially Quenched and Three Flavour ChPT \\
at Two Loops%
\thanks{Presented by JB.}%
}
\author{Johan Bijnens$^{a}$, Niclas Danielsson$^{a,b}$ and
Timo A. L\"ahde$^{c}$
\address{
$^a$ Dept. of Theor. Phys., Lund Univ.,
S\"olvegatan 14A, SE 22362 Lund, Sweden \\[2mm]
$^b$ Div. of Math. Phys., Lund Univ.,
S\"olvegatan 14A, SE 22362 Lund, Sweden \\[2mm]
$^c$ Dept. of Phys., Univ. of Washington, 
Seattle, WA 98195-1560, USA
}}
\maketitle
\vspace{-.5cm}
\begin{abstract}
A summary of recent progress in Chiral Perturbation Theory (ChPT) at the two-loop 
level is given. A short introduction to ChPT is included, along with an explanation 
of the usefulness of developing ChPT for partially quenched QCD. Further, our recent 
work in partially quenched ChPT is reviewed, and a few comments are given on older 
work in mesonic ChPT at the two-loop level. In particular, we quote the present 
best values for the low-energy constants of the $\mathcal{O}(p^4)$ chiral Lagrangian. 
\end{abstract}
  
\section{Introduction}
\label{introduction}

This talk describes some of the work done at two-loop order in mesonic
Chiral Perturbation Theory (ChPT) during the \mbox{EURODAPHNE I,II} and 
EURIDICE networks. A more extensive review of ChPT at this order
can be found in Ref.~\cite{Bijnensreview}. The aim of this talk is not to 
provide a full introduction or review of the two-loop work, but rather to 
concentrate on a few key issues. The outline of this talk is as follows: 
Sect.~\ref{chpt} gives a short introduction to ChPT, Sects.~\ref{pqqcd} 
and~\ref{pqchpt} discuss partial quenching, why it is thought to be useful enough to 
warrant two-loop calculations in partially quenched ChPT (PQChPT), and why
calculations in the PQChPT sector are much more challenging than those in
standard ChPT. Our own work in PQChPT is also briefly reviewed, and finally 
Sect.~\ref{standardchpt} gives a brief summary of other existing work at 
the two-loop order in ChPT.

\section{Chiral Perturbation Theory}
\label{chpt}

\begin{figure}[h!]
\begin{center}
\includegraphics{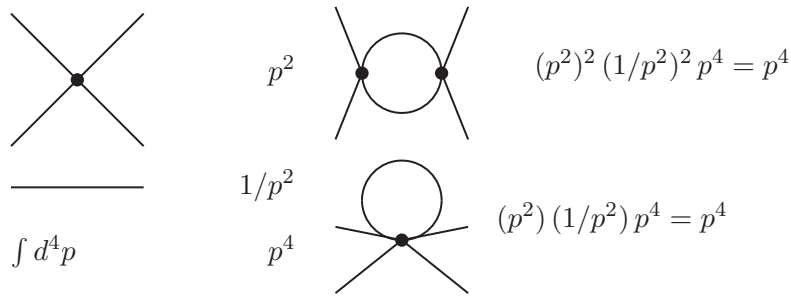}
\caption{Illustration of power counting in ChPT. On the left are shown: The
lowest order vertex, the meson propagator, a loop momentum integration and their 
respective powers of a generic momentum $p$. The examples on the right show two 
one-loop diagrams that count as $\mathcal{O}(p^4)$ as compared to $\mathcal{O}(p^2)$ 
for the lowest order.}
\label{figpower}
\end{center}
\end{figure}

Chiral Perturbation Theory was introduced in the papers by Weinberg, Gasser and
Leutwyler~\cite{Weinberg,GL1,GL2} which build on earlier work within current 
algebra and non-analytic higher order corrections. It should be noted that since 
a significant number of highly detailed introductions and lectures 
exist~\cite{CHPTlectures,website}, only a few of the main aspects are given here. 
The QCD Lagrangian
\begin{equation}
{\cal L}_{\mathrm{QCD}}^{} 
= -\frac{1}{4}G_{\mu\nu}^{}G^{\mu\nu}_{} + 
\hspace{-.2cm}\sum_{q=u,d,s}\hspace{-.2cm}
\left[
i\bar q_L^{} D\hskip-1.3ex/\, q_L^{} + 
i\bar q_R^{} D\hskip-1.3ex/\, q_R^{} - 
m_q^{}\left(\bar q_R^{} q_L^{} + \bar q_L^{} q_R^{} \right)
\right]
\label{lagqcd}
\end{equation}
is invariant under the chiral symmetry $SU(3)_L^{}\times SU(3)_R^{}$ when the 
masses of the up, down and strange quarks are set to zero. This symmetry is
expected to be spontaneously broken by the quark-antiquark vacuum expectation 
value
\begin{equation}
\langle \bar q q\rangle = \langle \bar q_L^{} q_R^{} + \bar q_R^{} q_L^{}\rangle
\ne 0
\end{equation}
to the diagonal subgroup $SU(3)_V^{}$. Since this involves the spontaneous
breaking of 8 generators of a global symmetry group, Goldstone's theorem requires 
the existence of 8 massless degrees of freedom {\em and} that their interactions 
vanish at zero momentum. ChPT is an effective field theory built on these eight 
massless particles which are identified with the pions, kaons and eta. This 
involves a long-distance expansion in momenta and quark masses. Such an expansion, 
called {\em power counting}, is possible because the interaction vanishes at 
zero momentum, and was worked out to all orders by Weinberg~\cite{Weinberg}. An 
example from ChPT is shown in Fig.~\ref{figpower}.

\vspace{-.5cm}

In order to perform the power counting expansion, higher order Lagrangians
need to be constructed. This has to be done in order to know the total number 
of parameters needed at a given order in the expansion. These parameters are 
referred to as low-energy constants (LECs). This classification was done at 
$\mathcal{O}(p^4)$ by Gasser and Leutwyler~\cite{GL1,GL2} and at $\mathcal{O}(p^6)$ 
in Ref.~\cite{BCE1}. The number of LECs needed for the partially quenched case was
determined in Ref.~\cite{BDL1,BDL2}. A summary of these results is given in
Table~\ref{tabparam}.

\begin{table}[ht!]
\begin{center}
\caption{The number of LECs (physical + contact terms) at the various
orders in mesonic ChPT.}
\label{tabparam}
\begin{tabular}{c|lc|lc|lc}
& \multicolumn{2}{c|}{2 flavour}
& \multicolumn{2}{c|}{3 flavour}
& \multicolumn{2}{c}{3+3 PQ} \\
\hline\hline
&&&&&& \vspace{-.3cm} \\
$\mathcal{O}(p^2)$ &
$F,B$ & 2 & $F_0^{},B_0^{}$ & 2 & $F_0^{},B_0^{}$ & 2 \\
$\mathcal{O}(p^4)$ &
$l_i^r,h_i^r$ & 7+3 & $L_i^r,H_i^r$ & 10+2 & $\hat L_i^r,\hat H_i^r$ & 11+2 \\
$\mathcal{O}(p^6)$ &
$c_i^r$ & 53+4 & $C_i^r$ & 90+4 & $K_i^r$ & 112+3 \\
\end{tabular}
\end{center}
\end{table}
\vspace{-.4cm}

The main problem here is to determine a minimal set of LECs. No simple and straightforward 
procedure is known for the determination of such a set. Only if all of the LECs can be 
separately determined from ``experiment'', or more generally from QCD Green's functions, 
can one be sure that the set of LECs is indeed minimal. Heat kernel methods allow to 
determine the divergence structure independently of Feynman diagram calculations. 
This, done at $\mathcal{O}(p^4)$ in~\cite{GL1,GL2} and at $\mathcal{O}(p^6)$ 
in~\cite{BCE2}, provides a very welcome check on actual two-loop calculations.

\section{Partially Quenched QCD}
\label{pqqcd}

One of the major applications of ChPT at present, and likely even more
so in the future, is the extrapolation of lattice QCD results to the physical 
values of the light $u,d$ quark masses. An overview of the many uses of ChPT
in lattice QCD can be found in the recent lectures by Sharpe~\cite{Sharpelectures}. 
The main emphasis here is on the partially quenched aspect, which can be implemented 
in lattice gauge theory. In order to extract observables, 
one typically evaluates a correlator, e.g. a two-point correlator to obtain masses
and decay constants. This correlator is evaluated in Euclidean space
via the path integral (or functional integral) formalism:
\begin{eqnarray}
\langle 0|
(\overline u\gamma_5^{} d)(x)(\overline d\gamma_5^{} u)(0)
|0\rangle
&=&
\dsp\int [dq][d\overline q][dG]\:
(\overline u\gamma_5^{} d)(x)(\overline d\gamma_5^{} u)(0)
\nonumber \\
&\times& \exp\left[{\dsp i\int d^4y\:\mathcal{L}_{\mathrm{QCD}}^{}}\right],
\end{eqnarray}
where the integral over the quarks and the anti-quarks can be performed
and one obtains, schematically,
\begin{eqnarray}
\label{lqcdint}
&& \hspace{-.5cm} \dsp\int [dq][d\overline q][dG]\:
(\overline u\gamma_5^{} d)(x)(\overline d\gamma_5^{} u)(0)\:
\exp\left[{\dsp i\int d^4y\:\mathcal{L}_{\mathrm{QCD}}^{}}\right]
\:\propto \\
&& \hspace{-.5cm} \int [dG] \:\underbrace{
\exp\left[-i\int d^4x\:\frac{G_{\mu\nu}G^{\mu\nu}}{4}\right]}_{\mbox{gluonic}}
\:\overbrace{
(D\hskip-0.6em/\,^u_G)^{-1}(x,0)(D\hskip-0.6em/\,^d_G)^{-1}(0,x)}^{\mbox{valence}}
\:\underbrace{
\det \left(D\hskip-0.6em/\,_G^{}\right)_{\mathrm{QCD}}^{}}_{\mbox{sea}},
\nonumber
\end{eqnarray}
where $D\hskip-0.6em/\,_G$ denotes the full Dirac operator with a specific gluon 
field configuration but including the quark masses. The remaining integral over 
all the gluon degrees of freedom in Eq.~(\ref{lqcdint}) is performed by 
importance sampling. The part labeled ``valence'' is connected to the external 
sources (hence the name), while the part labeled ``sea'' describes the effects 
of closed quark loops, not connected to any outside lines. Of course, gluons provide
couplings between all these fermion lines if we look at the functional
integral as a sum over Feynman diagrams.

One major problem is that the determinant labeled ``sea'' is extremely
CPU time consuming to evaluate. This has led to several approximations,
the most drastic of which is the quenched approximation, whereby the sea 
contribution is completely neglected and only the gluonic and valence ones 
retained. ``Unquenched'' means in this respect that the sea determinant is 
included. However, the high CPU time requirements make it difficult to vary 
the quark masses very much in this part, and changing quark masses in the 
part labelled ``valence'' is indeed computationally much cheaper. In partially
quenched simulations one thus varies the sea and valence quark masses 
independently of each other. There are good arguments in favour of this 
approach:
\begin{itemize}
\setlength{\parskip}{0cm}
\item It is clearly superior to the quenched approximation.
\item More systematic studies of the input parameters may be performed.
\item It turns out that some quantities can be extracted from different
observables in this way.
\item Unlike the quenched approximation, it is continuously 
connected to the QCD case.
\end{itemize}
However, a number of drawbacks need to be remembered:
\begin{itemize}
\setlength{\parskip}{0cm}
\item It is not QCD as soon as quark masses are different in the
valence and sea sectors.
\item It is not a {\it bona fide} Quantum Field Theory so the spin-statistics 
theorem and unitarity relations are not satisfied.
\end{itemize}
Especially the latter point might be important, since the derivation of ChPT from
QCD relies heavily on unitarity. Nonetheless, one expects that at least close to
the QCD case, PQQCD will have a low-energy effective theory similar to ChPT.

\section{Partially Quenched ChPT at Two Loops}
\label{pqchpt}

\begin{figure}[h!]
\begin{center}
\includegraphics{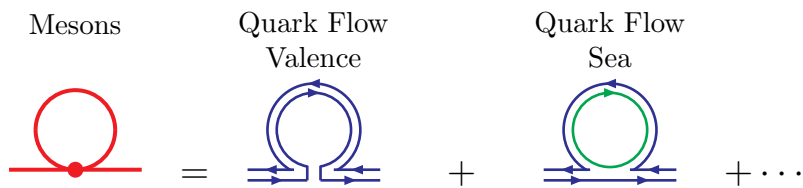}
\caption{The meson loop diagram on the left has different types of quark flow,
both valence and sea quark as indicated on the right.}
\label{figquark1}
\end{center}
\end{figure}

\begin{figure}[b!]
\begin{center}
\includegraphics{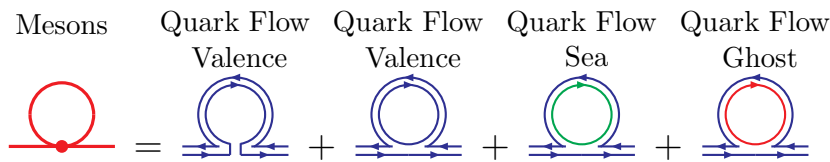}
\caption{The effect of adding ghost quarks to the different quark loops 
in the mesonic one loop diagram shown on the right.}
\label{figquark2}
\end{center}
\end{figure}

The central problem in PQChPT is thus to mimic, within ChPT, the effect of 
treating closed quark loops and quark lines differently. This problem is depicted
schematically in Fig.~\ref{figquark1}. In some of the early calculations, the 
quark flow was inferred directly from the flavour 
flow in the ChPT vertices. An alternative approach, often referred to as the 
supersymmetric method, is more systematic. A series of bosonic ghost quarks with
spin 1/2 is added to QCD. Due to the different statistics, these may cancel the 
effects of closed loops of valence quarks. This method is illustrated in 
Fig.~\ref{figquark2}.

\vspace{-.4cm}
The supersymmetric method
was originally introduced for the quenched case in Refs.~\cite{Morel,BG1,SharpeA}
and later extended to the partially quenched case, see 
Refs.~\cite{BG2,Sharpe1,Sharpe2} and references therein. Also, an instructive
discussion about ChPT in the partially quenched sector is given in Ref.~\cite{Sharpe1}.
For practical purposes, the QCD chiral symmetry may be replaced by a graded symmetry 
which is (assumed to be) spontaneously broken to its diagonal subgroup: 
\ba
SU(n_v^{}+n_s^{}|n_v^{})_L^{} \times 
SU(n_v^{}+n_s^{}|n_v^{})_R^{} \to 
SU(n_v^{}+n_s^{}|n_v^{})_V^{},
\ea
where $n_v^{}, n_s^{}$ denote the number of valence and sea quark flavours.
The ``Goldstone bosons'' now have both fermionic and bosonic character.
A large amount of work exists at one-loop order, see the references 
in~\cite{Sharpelectures}. One stumbling block at two-loop order 
was the determination of the divergence structure and Lagrangians.
Fortunately, it was realized~\cite{BDL1,BL1,BL2,BDL2} that the work 
of~\cite{BCE1,BCE2} could be taken over by formally replacing traces by
supertraces and the general number of flavours by the number of sea quarks.

\begin{figure}[h!]
\begin{center}
\includegraphics{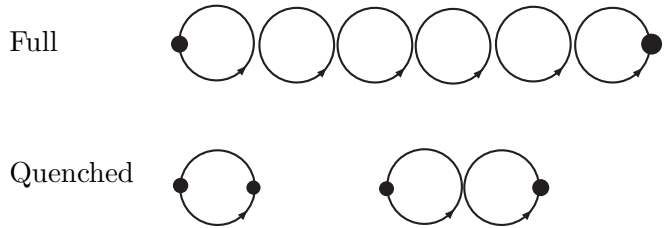}
\caption{Schematic resummation of quark loops.
Imagine gluons everywhere on top of the quark lines drawn.
Each bubble gives a lowest order meson propagator. 
The full resummation exponentiates and leads to a single pole.
For valence quarks only the loops shown at the bottom appear, leading to
double poles in the propagator.} 
\label{figquarkloop3}
\end{center}
\end{figure}

The final expressions at two-loop order are highly complex, which is due
in part to the larger number of independent quark masses, but mainly to the 
peculiarities of the flavour neutral mesons. These do not have a simple pole 
structure but consist instead of several terms including a double pole:
\be
\label{npropii}
\!\!-i\,G_{ij}^n (k) = \frac{\epsilon_j^{}}{k^2 \!-\! \chi_{ij}^{}}
- \frac{1}{n_{\mathrm{sea}}^{}}
\left[ \frac{R^d_i}{(k^2 \!-\! \chi_i^{})^2} 
+ \frac{R^c_i}{k^2 \!-\! \chi_i^{}}
+ \frac{R_{\eta ii}^\pi} {k^2 \!-\! \chi_\pi^{}} 
+ \frac{R_{\pi ii}^\eta}{k^2 \!-\! \chi_\eta^{}}
\right],
\ee
where the various $R$ coefficients consist of powers of ratios of
differences of quark masses~\cite{BG1,Sharpe1}. The presence of the many
ratios leads to the extremely long expressions. The double pole in 
Eq.~(\ref{npropii}) is due to the fact that PQChPT is not a full
field theory. Thus the valence quark loops cannot be resummed to all orders, 
as shown in Fig.~\ref{figquarkloop3}.

\vspace{-.4cm}
A large amount of work exists in PQChPT at two loops. The first calculation 
was the meson mass for the case of three sea quarks, with a common valence mass and a 
common sea quark mass~\cite{BDL1}. Since 
then the decay constants~\cite{BL1} and masses~\cite{BDL2} have been fully worked out. 
These quantities are also fully known for two sea quark flavours~\cite{BL2}. The 
formulas are rather lengthy and the number of parameters is also quite high.
Analytical programs have therefore been posted by the authors 
on the website~\cite{website}. Methods of dealing with the large number of parameters 
have been discussed extensively in Refs.~\cite{BL2} and~\cite{BDL2} for the two- and 
three-flavour cases, respectively. It should be emphasized again that the LECs of 
unquenched ChPT, and thus the full QCD results, are related to the partially
quenched LECs in a simple way via the Cayley-Hamilton relations of Ref.~\cite{BCE1}.
More recent work has focused on the neutral mass sector of PQChPT. It was shown 
in Ref.~\cite{Sharpe1} how the residue ${\cal D}$ of the double pole
\vspace{-.1cm}
\be
G^n_{ij}(k) =
\frac{-i{\cal Z}{\cal D}}{(k^2-M^2_{ch})^2} \:+\: \cdots
\ee
can be measured on the lattice and used at $\mathcal{O}(p^4)$ to extract
$L_7^r$, which is relevant for the $\eta$ mass. In Ref.~\cite{BD1} the self-energy 
resummation was worked out to all orders, and it was shown explicitly how to obtain 
the double pole and the structure of the full propagator from the
one-particle irreducible diagrams. It was also shown that all $\mathcal{O}(p^6)$ 
parameters relevant for the $\eta$ mass can be extracted using this method. The 
most recent work has focused on the inclusion of dynamical photons in the
partially quenched theory~\cite{BD2}.

\section{Standard ChPT at Two Loops}
\label{standardchpt}

The existing work at two-loop order in mesonic ChPT is very 
briefly reviewed here. A much more extensive review may be found in 
Ref.~\cite{Bijnensreview}. The oldest two-loop work in ChPT made use of 
dispersive techniques to calculate the non-analytical dependence on the kinematical
variables. This was done numerically~\cite{GM}
and analytically~\cite{CFU} for the pion vector and scalar form factors,
and fully analytically for $\pi\pi$ scattering in Ref.~\cite{Knechtpipi}. The 
first full two-loop calculations appeared somewhat later in the two-flavour 
sector, with $\gamma\gamma\to\pi^0\pi^0$~\cite{BGS} and $\gamma\gamma\to\pi^+\pi^-$,
$F_\pi$ and $m_\pi$~\cite{Buergi}. The process $\gamma\gamma\to\pi\pi$ was recently
recalculated in Ref.~\cite{GIS}. With $\pi\pi$ scattering \cite{BCEGS}, pion vector 
and scalar form factors~\cite{BCT} and the radiative decay
of the pion~\cite{BT1}, most processes of interest have now been worked out.

The earliest three-flavour work, on the vector two-point functions, was by Golowich 
and Kambor~\cite{GK1}, extended to all flavour cases in Refs.~\cite{DK,ABT1,Maltman1}. 
The first calculations with proper two-loop integrals 
were of the meson masses and decay constants, in Refs.~\cite{GK2,ABT1} 
and~\cite{ABT4}, including isospin violation. All scalar two point 
functions~\cite{Moussallam2,JBunpublished} and vacuum expectation values~\cite{ABT2}
are also known. More recent work covers the electromagnetic form factors~\cite{PS,BT2}, 
$K_{\ell3}^{}$~\cite{PS,BT3}, and scalar form-factors~\cite{BD}. Processes with more 
external legs include $K_{ell4}^{}$~\cite{ABT2}, kaon radiative decay~\cite{Geng},
$\pi\pi$~\cite{BDT} and $\pi K$~\cite{BDT2} scattering. The first 
results at finite volume have appeared recently~\cite{BG,CDH}.

A major problem in phenomenological applications of ChPT at two loops 
is to find enough experimental inputs to determine the 
$\mathcal{O}(p^6)$ parameters. In practice most of them have to be 
estimated, which is typically done along the lines of Ref.~\cite{EGPR} 
by saturating the LECs by resonance exchange. While this can be done at 
various levels of sophistication, most phenomenological applications 
have used a fairly simple extension of Ref.~\cite{EGPR},
see e.g. Refs.~\cite{BCEGS,ABT2,ABT4,BT2,BT3,BDT,BDT2}.

Most phenomenological applications rely on the work of Refs.~\cite{ABT2,ABT4}.
The fitting method and the inputs used are described in detail in 
Ref.~\cite{ABT4}, and the results can be found in Table~\ref{tabLi}, in the 
columns labeled ``fit 10''. These used the (at that time) most recent data 
of the BNL~E865 experiment as the main $K_{\ell4}^{}$ input. The change 
compared to a fit at $\mathcal{O}(p^4)$ is also given in Table~\ref{tabLi}.
These fits assume that the $1/N_c$ suppressed LECs $L_4^r$ and $L_6^r$ 
vanish at the scale $\mu=0.77$~GeV. On the other hand, ``Fit D'' of 
Ref.~\cite{BDT2} uses all the same inputs as ``fit 10'', in addition to 
the dispersive results on $\pi\pi$ and $\pi K$ scattering from Refs.~\cite{CGL}
and~\cite{BDM}. The convergence properties of some quantities are also given 
in Table~\ref{tabLi}. However, an update of the fit is in order, with
the new experimental results on $K_{\ell4}^{}$ and an improved treatment of 
the $\mathcal{O}(p^6)$ constants along the lines of Ref.\cite{BGLP,Eetal}.

\begin{table}[t!]
\begin{center}
\caption{The fitted optimal values of the $L_i^r$ at $\mathcal{O}(p^4)$ and
$\mathcal{O}(p^6)$, and convergence behviour of several quantities
for the different fits~\cite{ABT4,BDT2}.}
\label{tabLi}
\vspace{.4cm}
\begin{tabular}{c||c|c|c}
& fit 10, $\mathcal{O}(p^6)$ & fit 10, $\mathcal{O}(p^4)$ & fit D \\
\hline\hline
&&& \vspace{-.3cm} \\
$10^3 L_1^r$ & $0.43\pm0.12$ & $0.38$ & $0.44$ \\
$10^3 L_2^r$ & $0.73\pm0.12$ & $1.59$ & $0.69$ \\
$10^3 L_3^r$ & $-2.53\pm0.37$ & $-2.91$ &$-2.33$ \\
$10^3 L_4^r$ & $\equiv0$    & $\equiv 0$& $\equiv0.2$ \\
$10^3 L_5^r$ & $0.97\pm0.11$& $1.46$ & $0.88$ \\
$10^3 L_6^r$ & $\equiv0$    & $\equiv 0$& $\equiv0$ \\
$10^3 L_7^r$ & $-0.31\pm0.14$&$-0.49$ & $-0.28$ \\
$10^3 L_8^r$ & $0.60\pm0.18$ & $1.00$ & $0.54$ \\
&&& \vspace{-.3cm} \\
\hline
&&& \vspace{-.3cm} \\
$2B_0^{} \hat m/m_\pi^2$ & 0.736 & 0.991 & 0.958 \\
$m_\pi^2$: $\mathcal{O}(p^4)$, $\mathcal{O}(p^6)$
& 0.006, 0.258 & 0.009, $\equiv 0$ & $-$0.091, 0.133 \\
$m_K^2$: $\mathcal{O}(p^4)$, $\mathcal{O}(p^6)$
& 0.007, 0.306 & 0.075, $\equiv 0$ & $-$0.096, 0.201 \\
$m_\eta^2$: $\mathcal{O}(p^4)$, $\mathcal{O}(p^6)$
& $-$0.052, 0.318 & 0.013, $\equiv 0$ & $-$0.151, 0.197 \\
$m_u/m_d$ & $0.45 \pm 0.05$ & 0.52  & 0.50 \\
&&& \vspace{-.3cm} \\
\hline
&&& \vspace{-.3cm} \\
$F_0^{}$ [MeV] & 87.7 & 81.1 & 80.4 \\
$\frac{F_K}{F_\pi}$: $\mathcal{O}(p^4)$, $\mathcal{O}(p^6)$
& 0.169, 0.051 & 0.22, $\equiv 0$ & 0.159, 0.061
\end{tabular}
\vspace{-.6cm}
\end{center}
\end{table}

\section{Conclusions}

ChPT at two-loop order is by now a very well developed field, where a 
large number of two- and three-flavour calculations have been performed. 
The use of the partially quenched results will hopefully allow for many of
the $\mathcal{O}(p^6)$ LECs to be determined from Lattice QCD, thus removing a
major stumbling block in phenomenological applications.
 
\section*{Acknowledgments}

We would like to thank Giulia Pancheri for her many years of dedicated
work of running the networks EURODAPHNE I, II and
EURIDICE. It has been a very rewarding experience scientifically as
well as personally.

This work is supported in part by
the European Commission (EC) RTN Network
Grant No. MRTN-CT-2006-035482 (FLAVIAnet), 
the European Community-Research Infrastructure
Activity Contract No. RII3-CT-2004-506078 (HadronPhysics)
the Swedish Research Council and the U.S. Department of Energy under Grant No.
DE-FG02-97ER41014 (TL).

\end{document}